\documentclass[aps,prd,reprint,nofootinbib,showkeys,showpacs,superscriptaddress,longbibliography,floatfix]{revtex4-1}

\pdfoutput=1

\usepackage{amsmath}
\usepackage{amssymb}
\usepackage[american]{babel}
\usepackage{booktabs}
\usepackage{dcolumn}  
\usepackage[T1]{fontenc}  
\usepackage{graphicx}  
\usepackage[colorlinks]{hyperref}  
\usepackage[utf8]{inputenc}  
\usepackage{multirow}
\usepackage{txfonts}  
\usepackage{url}
\usepackage[dvipsnames]{xcolor}
\usepackage{float}

\newcolumntype{d}[1]{D{.}{.}{#1}}
\newcolumntype{v}[1]{D{,}{,\ }{#1}}

\makeatletter

\newcommand{\Rmnum}[1]{\expandafter\@slowromancap\romannumeral #1@}
\makeatother

\renewcommand {\arraystretch}{1.3}

\hypersetup{
	citecolor = blue,
	linkcolor = RoyalPurple,
	urlcolor = blue
}

\begin{document}

\title{Extended $\Lambda$CDM model and viscous dark energy: A bayesian analysis}

\author{W. J. C. da Silva}
\email{williamjouse@fisica.ufrn.br}
\affiliation{Universidade Federal do Rio Grande do Norte,
	Departamento de F\'{\i}sica, Natal - RN, 59072-970, Brasil}

\author{R. Silva}
\email{raimundosilva@fisica.ufrn.br}
\affiliation{Universidade Federal do Rio Grande do Norte,
	Departamento de F\'{\i}sica, Natal - RN, 59072-970, Brasil}
\affiliation{Universidade do Estado do Rio Grande do Norte, Departamento de F\'{\i}sica, Mossor\'o - RN, 59610-210, Brasil}


\date{\today}

\begin{abstract}
We propose an approach considering the nonextensive effects in the context of the Verlinde theory
in order to address an extended cosmological model in the context of viscous dark energy. Specifically, this model leads to a tiny perturbation
in the dynamics of the expansion of the universe through the generalized Friedmann equations so-called the extended $\Lambda$CDM model.
From the observational test standpoint, we make a Bayesian analysis of the models of bulk viscosity for dark energy which follows the Eckart theory of bulk viscosity. These models are investigated through the context of both models $\Lambda$CDM and extended $\Lambda$CDM. The Bayesian analysis is performed using the data of CMB Distance priors, Baryon Acoustic Oscillations Measurements, Cosmic Chronometers, and SNe Ia distance measurements.
\end{abstract}

\maketitle

\section{Introduction}\label{}

The accelerated expansion of the universe has widely been corroborated by the greater amount of observational data, such as type Ia Supernovae \cite{riess0, perlmutter}, Baryon Acoustic Oscillations (BAO) \cite{bao0} and Cosmic Microwave Background (CMB) anisotropies \cite{planck, planck2018}. These observations converge to the standard model, the $\Lambda$CDM model, where cosmological constant $\Lambda$ is responsible by acceleration of the universe and CDM refers to the Cold Dark Matter. Although this model has been confirmed as the standard cosmological model, a theoretical explanation of the physical mechanism responsible by cosmic acceleration has been a significant challenge in the modern cosmology \cite{weinberg89}. From the observational standpoint, there is a tension associated with the measurements of Hubble parameter at $z = 0$ by CMB anisotropies \cite{planck, planck2018}, and Cepheids and Supernovae \cite{riess, riess1, riess2, htension0}. The other reported tension is related to measurements of the growth of matter density fluctuations between late-time observations and CMB anisotropies (see more details in Ref. \cite{htension1}). There are different approaches to solve these puzzles. Typically, they can be divided into modified general relativity \cite{modifiedGR} and dark energy models \cite{DE}. The first case assumes modification in the standard general relativity based on some physical phenomena. The latter case proposes a new description for dark energy, or scalar field within the general relativity framework.

Another idea addressing the dark energy has focused on the fluid description, with the thermodynamics being the core of this scenario (see, e.g., \cite{thermogeral} and references therein). Many cosmological models, which are extensions of $\Lambda$CDM, have typically addressed dissipative process like the bulk viscosity in order to provide a thermodynamical framework \cite{eos}. More recently, by considering that the dark energy presents a bulk viscosity mechanism, the dark energy models have been analyzed in the context of fluid \cite{VDE0}, in the context scalar field \cite{VDE1,VDE2} and in the modified general relativity framework \cite{VDE3}.

On the other hand, the thermodynamics and its microscopic approach (statistical mechanics and kinetic theory) have been extended in
order to face the so-called complex systems \cite{tsallis-review}. By summarizing, the non-additive (or nonextensive) framework is based on
the parametrization of the entropy formula which depends on a free parameter $q$ (also called entropic parameter),
and provides the Boltzmann-Gibbs (BG) entropy in the additive limit $q\rightarrow 1$.  Specifically, the nonextensive framework is related with the Tsallis entropy, which for a classical non-degenerated gas system of point particles reads  (unless explicitly stated, in our units $k_B = c=1$)
\begin{equation}\label{Sq}
S_q\, = -\int f^{q}\ln_q f d^3p, \,\,\,\,\,
\end{equation}
where $q$ is quantifying the degree of nonadditivity of $S_q$, $f$ is the distribution of momentum and $\ln_q(f)$ is the nonadditive $q$-logarithmic function whose  inverse is the $q$-exponential. Both functions are defined by:
\begin{eqnarray}\label{E1}
\ln_q(f)=\,(1-q)^{-1}(f^{1-q}-1),\,\,\, (f>0),\\
e_q(f) = [1 + (1-q)f]^{\frac{1}{1-q}},\,\,\,e_q(\ln_qf)=f,
\end{eqnarray}
which reduce to the standard expressions in the limit $q \rightarrow 1$. The above formulas also imply that for a gas system composed by two subsystems (A,B), the kinetic Tsallis measure verifies $S_q(A + B)= S_q(A) + S_q(B) + (1-q)S_q(A)S_q(B)$. Hence, for $q=1$  the logarithm extensive measure associated to the GB approach is recovered.

In the context of cosmology, many connections have been investigated, e.g.  the entropic cosmology for a generalized black hole entropy \cite{komatsuQ}, black holes formation \cite{25,27}, the modified Friedmann equations from Verlinde theory \cite{28}, the role of the $q$-statistics on the light dark matter fermions \cite{supernovas}, the new perspective for the holographic dark energy \cite{DETsallis}. Indeed, there are many connections between cosmology and nonextensive framework (see, e.g., \cite{outras} and references therein).

A recent study addressed a connection between dissipative processes and nonextensive framework \cite{humberto,william}. The
principle behind this connection is based on the so-called Nonextensive/Dissipative Correspondence (\texttt{NexDC}), being
associated with the microscopic description of the fluid through the Tsallis distribution function \cite{osada}. Specifically, the \texttt{NexDC} has been
implemented to describe viscous dark matter \cite{humberto}. In addition, by using  the nonextensive effect in the Verlinde's
theory standpoint, a general model was proposed in order to investigate viscous dark matter \cite{william}.

In this paper, we particularly are interested in the investigation from models of the viscous dark energy, which consider first order deviations from equilibrium, i.e., the Eckart theory. From the background standpoint, we study these models by taking into the account the modified Friedmann equations based on the connections between the Tsallis statistics and the Verlinde's conjecture \cite{william,28}. Specifically, by using the models of viscous dark energy
\cite{VDE0,VDE1,velten,humberto,william}, we follow two different route, namely: i) By considering the standard dynamic ($\Lambda$CDM model), we make a
Bayesian analysis in order to investigate the bulk viscous models for dark energy with different
forms of the bulk viscous coefficient and ii) By using a general dynamic (extended $\Lambda$CDM), based on the nonextensive effects and Verlinde theory, we repeat the Bayesian analysis in order to investigate those models \cite{VDE0,VDE1,velten,humberto,william}.

The paper follows the sequence: in Sec. \ref{sec:2} we summarize the assumptions behind of the generalized Friedmann equations for bulk viscosity process. In Sec. \ref{sec:3} we introduce the viscous dark energy models \cite{VDE0,VDE1, velten,humberto, william} considering the extended $\Lambda$CDM model. In order to constrain parameters and compare models, in Sec. \ref{sec:4}, we make a Bayesian Analysis based on the data of CMB Distance priors, Baryon Acoustic Oscillations Measurements, Cosmic Chronometers, and SNe Ia distance measurements.  The main results and discussion concerning our approach for the viscous dark energy models are presented in Sec. \ref{sec:4}.

\section{Background and Assumptions}\label{sec:2}

Let us introduce a phenomenological approach by assuming an imperfect fluid. Furthermore, the dissipative process is related with an energy source of the FLRW universe. In this description, the momentum-energy tensor reads

\begin{equation}\label{eq9}
T_{T}^{\mu\nu} = T^{\mu\nu} + \Delta T^{\mu\nu},
\end{equation}
where $T_{T}^{\mu\nu}$ is the total momentum-energy tensor, $T^{\mu\nu}$ is momentum-energy tensor of perfect fluid and $\Delta T^{\mu\nu}$ represents dissipative process such as heat flux, anisotropic-stress and bulk viscosity. Here, we will consider a homogeneous, isotropic and flat universe, then only dissipative process allowed is the bulk viscosity  \cite{weinberg1971}. The simplest approach to treat bulk viscosity process is derived of the Eckart theory, which is a noncausal approach to dissipative phenomena. In this concern, the bulk viscosity pressure is given by \cite{eckart}
\begin{equation}\label{eq10}
\Delta T^{\mu\nu} =\Pi h^{\mu\nu},
\end{equation}
where, $h^{\mu \nu} = g^{\mu\nu} + u^{\mu}u^{\nu}$ is the usual projector onto the local rest space of $u^{\mu}$
(four-velocity) and $g^{\mu\nu}$ is the FLRW metric. $\Pi$ is the bulk viscous pressure, which depends on the bulk viscosity
coefficient and the Hubble parameter, $\Pi = -3\xi H$. Albeit this formalism has been widely used at background and
perturbative levels \cite{eos, VDE0, VDE1, VDE3}, its fundamental difficulty is related with its noncausal behavior, i.e,
since it admits dissipative signals with superluminal velocities \cite{maartens,israel1976,Kephart15,Li09,pavon91}.
A possible solution in order to face this difficult in the cosmological context, would be use a causal extension of the
Eckart framework, which is the so-called Israel-Stewart (IS) theory \cite{israel1976,maartens1996}. Issues on the fundament,
which have approached the problem of causality as well as the Ostrogradsky ghost  have been addressed by considering the IS
theory and Lagrangian formalism \cite{prd2016}, however, these issues are yet under debate \cite{Rezzola13}.  Indeed, IS
theory presented a better description than Eckart theory and Landau and Lifshitz theories, however like them, the common
behavior is associated to small deviation of equilibrium \cite{israel1976}. Recently, another connection with cosmology
has been proposed
through the full causal theory in the context of the acceleration of the universe \cite{mohana2017} and dark matter and
dark energy as a viscous single fluid \cite{piatella2011}. There are other connections between  the full causal theory and
cosmology (see, e.g., \cite{samuel17} and references therein). On the other hand, even though the Eckart theory has drawbacks
at fundamental level, it is the simplest than the IS theory, being widely used in order to investigate the accelerating
universe with the bulk viscous fluid (see, e.g. Refs.\cite{Fabris06,Kremer03,Cataldo05,Hu06,Athira15}). From the early
inflation standpoint, the Ref. \cite{hiscock1991} has shown that both the pathological Eckart theory and the truncated IS
theory provide inflation\footnote{From the cosmological perspective, the Ref. \cite{hiscock1991} has demonstrated that IS
approach converges to the Eckart’s theory, when the collision time-scale in the transport equation of fluid is zero, i.e.,
the bulk viscous model is necessarily noncausal and unstable \cite{maartens1996}.}. But the truncated version of IS theory
presented a constant relaxation time, being incorrect for an expanding universe. As we are investigating the viscous dark
energy, which is the component of the dark sector that provides the late-time acceleration of the expanding universe, we will
consider the models \cite{VDE0,VDE1,velten,humberto,william} which have used the Eckart formalism as a first order limit of
the IS theory with zero relation time. Moreover, this framework is a plausible approach to investigate the viscous dark energy
since the physical effect occurs on the pressure, being a tiny perturbation of the standard $\Lambda$CDM model.

Now, by choosing a reference frame in which the hydrodynamics four-velocity $u^{\mu}$ is unitary, $u^{\mu}u_{\mu} = -1$, and replacing the Eq.(\ref{eq10}) into Eq.(\ref{eq9}), we obtain

\begin{equation}\label{eq11}
T_{T}^{\mu\nu} = (\rho + P_{\text{eff}})u^{\mu}u^{\nu} + P_{\text{eff}}g^{\mu\nu},
\end{equation}
where $\rho $ is the energy density, $P_{\text{eff}} = p_{k} + \Pi$, where $p_k$ is the kinetic pressure
(equilibrium pressure) and $\Pi = -3\xi H$.

By considering the conservation equation $\nabla_\mu T^{\mu}_{\nu} = 0$ in Eq. (\ref{eq11}) one finds
\begin{equation}\label{eq12}
\dot{\rho} + 3H(\rho + p_k) - 9H^{2}\xi = 0.
\end{equation}
This is the energy conservation equation for viscous fluid. The functional form of $\xi$ is fundamental to the dynamics of the model.

In order to compare some models of viscous dark energy through the Bayesian
Analysis, let us consider the so called the extended Friedmann equations \cite{28,william}\footnote{In the Verlinde's conjecture \cite{19}, the gravity is explained as an entropic force caused by changes in the information associated with the positions of particles. The assumption of the entropic force, together with the Unruh temperature, provides the derivation of the second law of Newton. Moreover, by considering the holographic principle and the equipartition law of energy, this approach leads to Newton's law of gravitation. These ideas have been used in order to propose a thermodynamic derivation of Einstein equations \cite{jacobson1995}. In this regards, it was demonstrated in Ref. \cite{28} through arguments of the Refs. \cite{19,cai} that one modification in the field equations can be obtained simply by assuming the nonextensive equipartition law of energy. From the mathematical standpoint, this extended approach leads to an effective gravitational constant, i.e., $G \rightarrow G_q =  \frac{5-3q}{2}G$ \cite{28,william}.}
\begin{equation}\label{extended-friedmann}
H^{2} = \frac{8\pi G}{3}\rho\left(\frac{5 - 3q}{2}\right) - \frac{k}{a^2},
\end{equation} and
\begin{equation}\label{27}
\frac{\ddot{a}}{a} = -\frac{4\pi}{3}\left(\frac{5 - 3q}{2}\right)G(\rho +3p),
\end{equation}
where $H = \frac{\dot{a}}{a}$ is the Hubble parameter, $q$ is the nonextensive parameter, $\rho$ is the total energy density,
$p$ is the pressure of the fluid and, $k$ represents the spatial curvature. Here, we will consider the flat and
non-flat Universe $k=0,\pm 1$, and the extended expression (\ref{27}) in order to investigate different models of viscous
dark energy.

\section{Viscous dark energy}\label{sec:3}
In this section, we consider the Friedmann equations obtained in the previous one and the dark energy as a fluid with bulk
viscosity process. The main contributions to the total momentum-energy tensor of the cosmic fluid are the radiation,
the baryonic matter, cold dark matter and the viscous dark energy. The radiation, baryons and dark matter are assumed to have
the usual properties of perfect fluids. As each component of the cosmic fluid is individually conserved, we obtain

\begin{equation}\label{fluid-conservation}
\dot{\rho_\text{i}} + 3H(\rho_\text{i}+p_{\text{i}})=0,
\end{equation}
where $\text{i}$ is related to the radiation (r), the baryonic matter (b), cold dark matter (dm). By considering the Eq. (\ref{eq12}), the energy conservation of viscous dark energy is given by

\begin{equation}\label{viscous-conservation}
\dot{\rho}_{\text{de}} + 3H(\rho_{\text{de}} + \tilde{p}_{\text{de}}) = 0,
\end{equation}
where $\rho_{\text{de}}$ the energy density of viscous dark energy, $\tilde{p}_{\text{de}} = p_k + \Pi$ is the effective
pressure, $p_k$ is the equilibrium pressure, $\Pi = -3\xi H$, bulk viscosity pressure and $\xi$ is
the bulk viscosity coefficient. The choice of bulk viscosity coefficient $\xi$ generates different viscous
dark energy models. The general case, $\xi$ is not constant, and in the literature there are different approaches to
determining how bulk viscosity evolves. We consider three different bulk viscosity functions in our analysis: (i) bulk
viscosity being proportional to the Hubble parameter, $\xi = \xi_0 H$; (ii) bulk viscosity proportional to energy density
and inversely proportional to Hubble parameter; (iii) the usual ansatz for the bulk viscosity, a function for thermodynamical
state, i.e., energy density of the fluid, in the case $\xi = \xi(\rho_{\text{de}})$.

\subsection{Model \Rmnum{1}}
The first model analyzed is the bulk viscosity proportional to the Hubble parameter, i.e, from the Friedmann equation, the bulk viscosity is proportional to the square root of the total energy density. This dependency allows us to consider that bulk viscosity is a function of all the other cosmological fluids. The model was studied in Ref. \cite{VDE0, avelino2013}, with the ansatz for bulk viscosity evolution given by

\begin{equation}
\xi = \xi_{0}H,
\end{equation}
and, the effective pressure reads

\begin{equation}\label{1.16}
\tilde{p}_{\text{de}} = p_{\text{de}} - 3H^{2}\xi_0,
\end{equation}
where $\xi_0$ is the current value for bulk viscosity and $p_{\text{de}} = \omega \rho_{\text{de}}$. Firstly, we consider parameter of equation of state, $\omega = -1$, consequently, the effective pressure is $\tilde{p}_{\text{\text{de}}} = -\rho_{\text{\text{de}}} - 3\xi_0 H^2$. We call this Model \Rmnum{1}a. The second case, we consider $\omega$ as free parameter, this model is called Model \Rmnum{1}b. Afterwards, combining the Eqs. (\ref{extended-friedmann}), (\ref{fluid-conservation}), (\ref{viscous-conservation}) and (\ref{1.16}), the Friedmann equation for Model \Rmnum{1}a is given by

\begin{equation}\label{hubble1}
\begin{aligned}
\frac{H^{2}}{H_{0}^2}  = & \left(\frac{5 - 3q}{2}\right) \Bigg[\Omega_{\text{b}}(1 + z)^{3} + \Omega_{\text{r}}(1 + z)^{4}+ \frac{\Omega_{\text{dm}}}{1 + \tilde{\xi}}(1 + z)^3 \\ & + \Bigg(1 - \frac{\Omega_{\text{dm}}}{1 +  \tilde{\xi}}\Bigg)(1 + z)^{-3\tilde{\xi}}\Bigg],
\end{aligned}	
\end{equation}
where $z$ is the redshift, $\Omega_{\text{dm}}$ is the matter density parameter today and $\tilde{\xi}$ is dimensionless bulk viscosity. For Model \Rmnum{1}b, the Friedmann equations reads

\begin{equation}\label{hubble2}
\begin{aligned}
\frac{H^{2}}{H_{0}^2} = &  \left(\frac{5 - 3q}{2}\right) \Bigg[\Omega_{\text{b}}(1 + z)^{3} + \Omega_{\text{r}}(1 + z)^{4} + \frac{\omega\Omega_{\text{dm}}}{\omega -  \tilde{\xi}}(1 + z)^3 \\ & + \Bigg(1 - \frac{\omega\Omega_{\text{dm}}}{\omega -  \tilde{\xi}}\Bigg)(1 + z)^{3(1 + \omega -  \tilde{\xi})}\Bigg].
\end{aligned}	
\end{equation}
Also, we consider bulk viscosity effects on the non-flat Universe. To make this, we add curvature density parameter evolution in the Model \Rmnum{1}a, and we name Model \Rmnum{1}c. Friedmann equation for this model is given by

\begin{equation}\label{hubble3}
\begin{aligned}
\frac{H^{2}}{H_{0}^2} & = \left(\frac{5 - 3q}{2}\right) \Bigg[\Omega_{\text{b}}(1 + z)^{3} + \Omega_{\text{r}}(1 + z)^{4} + \frac{\Omega_{\text{dm}}}{1 -  \tilde{\xi}}(1 + z)^3 \\ & + \frac{2\Omega_k }{2 + 3 \tilde{\xi}}(1 + z)^{2} + \Bigg(1 - \frac{2\Omega_k}{2 + 3 \tilde{\xi}} - \frac{\Omega_{\text{dm}}}{1 +  \tilde{\xi}}\Bigg)(1 + z)^{-3 \tilde{\xi}}\Bigg],
\end{aligned}	
\end{equation}
where $\Omega_{k}$ is the today curvature density parameter. The dimensionless bulk viscosity parameter is defined by
\begin{equation}
 \tilde{\xi} = \frac{8\pi G \xi_{0}}{H_0},
\end{equation}
is valid for all models.
\subsection{Model \Rmnum{2}}

Another interesting functional form for bulk viscosity is a ratio between corresponding energy density and expansion rate given by \cite{VDE1}

\begin{equation}
	\xi = 3\xi_0\frac{\sqrt{\rho_{\text{de}}}}{H},
\end{equation}
where $\xi_{0}$ is the present-day bulk viscosity and $\rho_{\text{de}}$, dark energy density. The effective pressure for this model is

\begin{equation}\label{p5}
	\tilde{p}_{\text{de}} = -\rho_{\text{de}} - 3\xi_0\sqrt{\rho_{\text{de}}}.
\end{equation}
For this ansatz, bulk viscosity of dark energy is insignificant in early Universe (when dark matter dominates). Any one way, the bulk viscosity increases in late Universe \cite{VDE1}.

From Eqs. (\ref{extended-friedmann}), (\ref{fluid-conservation}), (\ref{viscous-conservation}) and and (\ref{p5}) in the flat Universe, the evolution of the Friedmann equation for bulk viscosity model is given by

\begin{equation}\label{hubble4}
\begin{aligned}
\frac{H^{2}}{H_{0}^2} = & \left(\frac{5 - 3q}{2}\right) \Bigg[\Omega_{\text{b}}(1 + z)^{3} + \Omega_{\text{r}}(1 + z)^{4} +  \Omega_{\text{dm}}(1+z)^{3} \\ & + \Omega_{\text{de}}\Bigg(1 - \frac{9\tilde{\xi}}{2\sqrt{\Omega_{\text{de}}}}\ln (1 + z)\Bigg)^2\Bigg],
\end{aligned}
\end{equation}
where $\tilde{\xi}$ is the dimensionless bulk viscosity coefficient defined by
\begin{equation}
	 \tilde{\xi} = \sqrt{\frac{8\pi G}{3H^2_0}}\xi_{0}.
\end{equation}
We use the normalization condition $\Omega_{\text{de}} = \frac{2}{5-3q} - \Omega_{\text{b}} - \Omega_{\text{dm}} - \Omega_{\text{r}}$.

\subsection{Model \Rmnum{3}}

The last model of bulk viscosity considered in this work was studied by Refs. \cite{velten, humberto, william} in the context of viscous dark matter. Thus, assuming that the bulk viscosity is given by

\begin{equation}\label{xi-model3}
\xi = \xi_{0}\Bigg(\frac{\rho_{\text{de}}}{\rho_{\text{de}0}}\Bigg)^{\alpha},
\end{equation}
where $\xi_{0}$ is the current value for bulk viscosity, $\rho_{\text{de}0}$ is the density of the viscous dark energy today and $\alpha$ is constant. We can set $\alpha$ in two values, $\alpha = 0$ and $\alpha = -1/2$, for to alleviate the integrated Sachs-Wolfe effect \cite{velten}. Then, we consider only  $\alpha = 0$, then, the effective pressure for this model

\begin{equation}\label{1.9}
\tilde{p}_{\text{de}} =  p_{\text{de}} - 3H\xi_0,
\end{equation}
where $p_{\text{de}} = \omega \rho_{\text{de}}$ and $\xi_0$ is a constant parameter. The value of $\alpha = 0$ has a physical interpretation, means a constant bulk viscosity coefficient.

The Hubble expansion rate $H$ is given in terms of the energy densities $\Omega_i$ where the subscript $i$ corresponds to each fluid, i.e., dark matter, dark energy, radiation
\begin{equation}
	\begin{aligned}\label{hubble5}
	\frac{H^{2}}{H_{0}^{2}} = & \left(\frac{5 - 3q}{2}\right) \Big[\Omega_{\text{b}}(1 + z)^{3} + \Omega_{\text{r}}(1 + z)^{4} \\ & + \Omega_{\text{dm}}(1 + z)^{3} + \Omega_{\text{de}}(z)\Big],
	\end{aligned}
\end{equation}
where  $\Omega_{\text{b}}$, $\Omega_{\text{r}}$, $\Omega_{\text{dm}}$ are baryonic, radiation and dark matter density parameter today, respectively. In goal to determinate functional form of $\Omega_{\text{de}}$, we have to solve conservation equation (\ref{viscous-conservation}) with the effective pressure, Eq. (\ref{1.9}), then

\begin{equation}\label{1.12}
\begin{aligned}
\frac{d \Omega_{\text{de}}}{dz} = &\frac{3\Omega_{\text{de}}}{1 + z}(1 + \omega) -  \frac{\tilde{\xi}}{1 + z}\Big\{\left(\frac{5 - 3q}{2}\right) \Big[\Omega_{\text{b}}(1 + z)^{3} \\ & + \Omega_{\text{r}}(1 + z)^{4}+ \Omega_{\text{dm}}(1 + z)^{3} + \Omega_{\text{de}}(z)\Big]\Big\}^{1/2},
\end{aligned}	
\end{equation}
where we define a dimensionless bulk viscosity parameter as

\begin{equation}
\tilde{\xi} = \frac{24 \pi G \xi_{0}}{H_{0}},
\end{equation}
in which is valid for Eq. (\ref{1.12}). The initial condition for the differential equation is $\Omega_{\text{de}}(0) = \frac{2}{5-3q} - \Omega_{\text{b}} - \Omega_{\text{dm}} - \Omega_{\text{r}}$.

In the next sections we will present the cosmological data to do a Bayesian analysis of these viscous dark energy models in two ways: the first analysis will be done by fixing the value of $q = 1$, that is, without nonextensivity and in the second moment, we will consider the parameter $q$ as free in the analysis.

\section{Data Constraints and Bayesian Analysis}\label{sec:4}
In this section, we will present the data and technique used in this work. To constrain parameters and compare models, we perform Bayesian Analysis based on the presented data. Recently, Bayesian Analysis has been extensively used to constraint and compare cosmological models \cite{bethoven, simony, maria, uendert, antonella, william}. In our analysis, we consider CMB Distance priors derived from Planck 2015,  the eight baryon acoustic oscillations measurements \cite{wigglez, bao1,bao2,bao3,bao4}, 24 cosmic chronometers measurements from Ref. \cite{moresco} and $1048$ SNe Ia distance measurements of the Pan-STARRS (Pantheon) dataset \cite{scolnic}.

The $\Lambda$CDM model is assumed as reference model and is parameterized with following set of cosmological parameters: the dimensionless Hubble constant $h$, the baryon density parameter, $\Omega_{b}$, the cold dark matter density parameter $\Omega_{\text{dm}}$. The parameters of the other models are listed in the Table \ref{tab:prior} together with their priors. We choose uniform priors on baryon parameter $\Omega_b$ and cold dark matter parameter $\Omega_{dm}$. For dimensionless Hubble parameter $h$ we consider a range $10$ times wider than value obtained in Ref. \cite{riess}. For curvature parameter $\Omega_k$ and $\omega$, we adopt $1\sigma$ values reported by Planck Results \cite{planck} and uniform prior, respectively. The prior for bulk viscosity is based in recent results \cite{humberto, velten, william}. For nonextesivity parameter $q$ we assume the values that agree with Friedmann equation.  We fix $\Omega_{\gamma} h^2 = 2.469 \times 10^{-5}$ \cite{komatsu}, $\Omega_rh^2 = 1.698\Omega_{\gamma}$ \cite{mangano}.

The most important quantity for Bayesian model comparison is the Bayesian evidence, or marginal likelihood, and is obtained here by implementing the \textsf{PyMultiNest} \cite{johannes}, a Python interface for \textsf{MultiNest} \cite{feroz}, the Bayesian tool based on the nested sampling \cite{skilling} in which calculates the evidence, but still allows constrain parameters with consequence. We plot the results using \textsf{GetDist} \cite{getdist}.

We following the standard description (see Refs. \cite{trotta,antonella,bethoven,william}), the posterior distribution $P(\Theta|D, M)$ is given by
\begin{equation}\label{bayes}
P(\Theta|D, M) = \frac{\mathcal{L}(D|\Theta, M) \pi(\Theta|M)}{\mathcal{E}(D|M)},
\end{equation}
where $\mathcal{L}(D|\Theta, M)$, $\pi(\Theta|M)$ and $\mathcal{E}(D|M)$, the likelihood, the prior and Bayesian evidence with $\Theta$ denotes the parameters set, $D$ the cosmological data and $M$ the model. The evidence can be written in the continuous parameter space $\Omega$ as

\begin{equation}\label{evidence}
\mathcal{E} = \int_\Omega \mathcal{L}(D|\Theta, M) \pi(\Theta|M) d\Theta.
\end{equation}

In order to compare two models, $M_i$ and $M_j$, we compute the ratio of the posterior probabilities, given by \cite{trotta}

\begin{equation}
\frac{P(M_i|D)}{P(M_j|D)} = B_{ij}\frac{P(M_i)}{P(M_j)},
\end{equation}
where $B_{ij}$ is known as the Bayes factor, defined as

\begin{equation}\label{bayes_factor}
B_{ij} = \frac{\mathcal{E}_i}{\mathcal{E}_j}.
\end{equation}
The Bayes factor of the model $i$ relative to the model $j$ (here, we assumed to be the $\Lambda$CDM model). It is emphasized pointing out that the Bayesian evidence rewards models that balance the quality of fit and complexity \cite{liddle}. Indeed, the larger the number of free parameter, not required by the data, the penalization of the model will be greater than the other. The usual interpretation of the Bayes factor is related to Jeffreys' scale. We use an alternative version of Jeffreys' scale suggested in Ref. \cite{trotta}.

\begin{table}[H]
	\renewcommand{\arraystretch}{1.5}
	\renewcommand{\tabcolsep}{0.2cm}
	\centering
	\caption{The table shows the priors distribution used is this work.}
	\begin{ruledtabular}
		\begin{tabular}{cccl}
			Parameter      & Model & Prior 						  & Reference \\ \colrule
			$h$			   & All   & $\mathcal{U}(0.5584,0.9064)$ & \cite{riess} \\
			$\Omega_b$	   & All   & $\mathcal{U}(0.0005,0.1)$    &  -   \\
			$\Omega_{dm}$  & All   & $\mathcal{U}(0.001, 0.99)$   &  -   \\
			$\Omega_{k} $  & Model \textsf{Ic}  & $\mathcal{N}(-0.05, 0.05)$   & \cite{planck}\\
			$ \omega $     & $\omega$CDM, Model \textsf{Ib} & $\mathcal{U}(-2.0, 0.0)$ & -  \\
			$ \tilde{\xi}$     & All except $\Lambda$CDM& $\mathcal{N}(0.0, 0.1)$ & \cite{humberto, velten, william} \\
			$q$ & All except $\Lambda$CDM &$\mathcal{U}(0.8, 1.4)$ & - \\
		\end{tabular}
	\label{tab:prior}
	\end{ruledtabular}
	
\end{table}

\subsection{Baryon Acoustic Oscillations Measurements}
The interaction between gravitational force and primordial relativistic plasma generates acoustic oscillations at the recombination epoch, which leave their signature in every epoch of the Universe. The measurements of BAO provide an independent standard ruler to constrain cosmological models.

The BAO measurements are given in terms of angular scale and the redshift separation, this is obtained from the calculation of the spherical average of the BAO scale measurement, and it is given by \cite{eisenstein, eisenstein2}

\begin{equation}
	d_z = \frac{r_s(z_{\text{drag}})}{D_V(z)},
\end{equation}
in which $D_V(z)$ is volume-averaged distance given by

\begin{equation}
	D_V(z) = \Bigg[(1+z)^2D_A(z)^2\frac{cz}{H(z)}\Bigg]^{1/3},
\end{equation}
where $c$ is the speed of light and $D_A$ is the angular diameter distance given by

\begin{equation}
	D_A(z) = \frac{c}{1+z}\int_{0}^{z} \frac{dz}{H(z)}.
\end{equation}
$r_s(z_{\text{drag}})$ is the comoving size of the sound horizon calculated in redshift at the drag epoch defined by

\begin{equation}
	r_s(z_{drag}) = \int_{z_{drag}}^{\infty} \frac{c_s dz}{H(z)},
\end{equation}
in which $c_s(z) = \frac{c}{\sqrt{3(1 + \mathcal{R})}}$ is the sound speed of the photon-baryon fluid and
$\mathcal{R} = \frac{3}{4}\frac{\Omega_b}{\Omega_r}\frac{1}{1 + z}$. We consider the redshift at the drag epoch $z_{\text{drag}}$ given by \cite{eisenstein2}

\begin{equation}
 	z_{\text{drag}} = \frac{1291(\Omega_{\text{m}}h^2)^{0.251}}{1 + 0.659(\Omega_{\text{m}}h^2)^{0.828}}\Big[1 + b_{1}(\Omega_{\text{m}}h^2)^{b_{2}}\Big],
\end{equation}
where $b_1 = 0.313(\Omega_{\text{m}}h^2)^{-0.419}\Big[1 + 0.607(\Omega_{\text{m}}h^2)^{-0.674}\Big]$, $b_{2} = 0.238(\Omega_{\text{m}}h^2)^{0.223}$.

In this analysis we consider the BAO measurements from diverse surveys, see the Table \ref{tab:BAO}. Furthermore, we also include three measurements from WiggleZ Survey \cite{wigglez}: $d_z(z=0.44) = 0.073$, $d_z(z=0.6) = 0.0726$, and $d_z(z=0.73) = 0.0592$. These measurements are correlated by following inverse covariance matrix

\begin{equation}\label{C_BAO}
C^{-1} =
\begin{pmatrix}
1040.3 & -807.5  & 336.8 \\
-807.5 & 3720.3  & -1551.9 \\
336.8  & -1551.9 & 2914.9
\end{pmatrix} \,.
\end{equation}

\begin{table}[H]
	\centering
	\medskip
	\centering
	\renewcommand{\arraystretch}{1.5}
	\renewcommand{\tabcolsep}{0.2cm}
	\caption{\label{tab:BAO}BAO distance measurements for each survey considered.}
	\begin{ruledtabular}
		\begin{tabular}{l c c c}
			Survey     & \multicolumn{1}{c}{$z$} & \multicolumn{1}{c}{$d_z(z)$} & Ref. \\
			\colrule
			6dFGS      & $0.106$ & $0.3360 \pm 0.0150$ & \cite{bao1} \\
			MGS        & $0.15$  & $0.2239 \pm 0.0084$ & \cite{bao2} \\
			BOSS LOWZ  & $0.32$  & $0.1181 \pm 0.0024$ & \cite{bao4} \\
			SDSS(R)    & $0.35$  & $0.1126 \pm 0.0022$ & \cite{bao3} \\
			BOSS CMASS & $0.57$  & $0.0726 \pm 0.0007$ & \cite{bao4} \\
		\end{tabular}
	\end{ruledtabular}
\end{table}

For the WiggleZ data, the chi-squared function is

\begin{equation}
\chi_\text{WiggleZ}^2 = \textbf{D}^T \textbf{C}^{-1} \textbf{D},
\end{equation}
where $ \textbf{D} = \mathbf{d}_{z}^{\,\text{obs}} - \mathbf{d}_z^{\,\text{mod}}$ and $\textbf{C}^{-1}$ is the covariance
matrix given by Eq. (\ref{C_BAO}).

The chi-squared function related with each survey is given by

\begin{equation}
\chi_\text{Survey}^2 = \left[\frac{d_z^{\,\text{obs}}(z) - d_z^{\,\text{mod}}(z)}{\sigma_\text{Survey}} \right]^2,
\end{equation}
where $d_{z}^{\,\text{obs}}$ is the observed ratio value, $d_{z}^{\,\text{mod}}$ is theoretical ratio value and $\sigma$ is the uncertainties in the measurements for each data point.

Then, the BAO $\chi^2$ function contribution is defined as
\begin{equation}
\chi_\text{BAO}^2 = \chi_\text{WiggleZ}^2 + \chi_\text{Survey}^2.
\end{equation}

\subsection{CMB Distance Priors}
CMB distance priors can be derived from data, such as \textit{Planck collaboration} or WMAP from the full Boltzmann analysis of CMB data. In Refs. \cite{cmb1, cmb2, cmb3}, they discussed the possibility to compress CMB likelihood in few numbers: CMB shift parameter $\mathcal{R}$ \cite{cmb4}, the angular scale of the sound horizon at last scattering $\ell_A$, they are important to deal with the late-time expansion history, and baryon density today $\Omega_{b}h^2$, it is important to study the late-time Universe but not sensitive to the cosmological models.

CMB shift parameter is defined as

\begin{equation}\label{R}
	\mathcal{R} = \sqrt{\Omega_{m}H^2_0}r(z_\star)/c,
\end{equation}
where $r(z_\star) = \frac{c}{H_0}\int_{0}^{z} \frac{dz}{E(z)} $ and angular scale of the sound horizon at last scattering

\begin{equation}\label{la}
	\ell_A = \pi r(z_\star)/r_s(z_\star),
\end{equation}
where $r_s(z_\star)$ is comoving size of the sound horizon calculated in the redshift of decoupling epoch given by \cite{cmb5}

\begin{equation}
	z_\star = 1048[1 + 0.00124(\Omega_bh^2)^{-0.738}][1 + g_1(\Omega_{m}h^2)^{g_2}],
\end{equation}
where
\begin{equation}
	g_1 = \frac{0.0783\Omega_{b}^{-0.238}}{1 + 39.5(\Omega_{b}h^2)^{0.763}},
\end{equation}
\begin{equation}
	g_2 = \frac{0.560}{1 + 21.1(\Omega_{b}h^2)^{1.81}}.
\end{equation}

Then, the $\chi^2$ function of the CMB prior is defined as

\begin{equation}
	\chi^{2}_{CMB} = \textbf{X}^{T}_{CMB}\cdot \textbf{C}^{-1}_{CMB} \cdot\textbf{X}_{CMB},
\end{equation}
where $ \textbf{X}_{CMB} = (\mathcal{R}(z_\star), \ell_A(z_\star), \Omega_b h^2) - (\mathcal{R}^{obs}, \ell_A^{obs}, \Omega_b h^{2\,obs})$ with $\mathcal{R}^{\,obs} = 1.7488$, $\ell_A^{\,obs} = 301.76$, $\Omega_b h^{2\,obs} = 0.02228$ and covariance matrix $\textbf{C}$ from Planck Results \cite{planck}.

\subsection{Cosmic Chronometers}
Another analysis considered in this work are the cosmic chronometers obtained through the differential age method. The cosmic chronometer is a method to determine the Hubble parameter values at different redshifts taking the relative age of passively evolving galaxies \cite{h1, h2, h3}. The method calculates $dz/dt$ and hence the Hubble parameter is given by
\begin{equation}\label{H}
	H(z) = - \frac{1}{1+z}\frac{dz}{dt}.
\end{equation}
Here, the theoretical values of $H(z)$ are given by Eqs. (\ref{hubble1}), (\ref{hubble2}), (\ref{hubble3}), (\ref{hubble4}), (\ref{hubble5}). The measurement of $dz$ is obtained through spectroscopic data with high accuracy, then for a precise measurement of the Hubble parameter, it is necessary to measure the differential age evolution $dt$ of such galaxies, and hence cosmic chronometers are considered to be model independent. A  detailed  description  about  the  cosmic chronometer method can be found in Ref. \cite{h4,riess}. Here we use the 24 measurements of the Hubble parameter in the redshift interval $0.1 < z < 1.2$, which are listed in Table \ref{tab:h_clocks} \cite{moresco}. The choose of this measures is motivate by the following argument,the expansion history data of the Universe might no be smooth outside the quoted redshift range \cite{h4}.

\begin{table}[ht!]
	\renewcommand{\arraystretch}{1.5}
	\renewcommand{\tabcolsep}{0.2cm}
	\centering
	\medskip
	\caption{Estimated values of $H(z)$ obtained using the differential age method.}
	\begin{tabular}{cc|cc|cc}
		\hline
		\hline
		\multicolumn{1}{c}{$z$} & \multicolumn{1}{c}{$H(z)$} & \multicolumn{1}{c}{$z$} & \multicolumn{1}{c}{$H(z)$} & \multicolumn{1}{c}{$z$} & \multicolumn{1}{c}{$H(z)$}
		\\ \hline
		$0.07   $ & $ 69    \pm 19.6 $ &  $0.28   $ & $ 88.8  \pm 36.6 $ & $0.48   $ & $ 97    \pm 62   $ \\
		$0.09   $ & $ 69    \pm 12   $ &  $0.352  $ & $ 83    \pm 14   $ & $0.593  $ & $ 104   \pm 13   $ \\
		$0.12   $ & $ 68.6  \pm 26.2 $ &  $0.3802 $ & $ 83    \pm 13.5 $ & $0.68   $ & $ 92    \pm 8    $ \\
		$0.17   $ & $ 83    \pm 8    $ &  $0.4    $ & $ 95    \pm 17   $ & $0.781  $ & $ 105   \pm 12   $ \\
		$0.179  $ & $ 75    \pm 4    $ &  $0.4004 $ & $ 77    \pm 10.2 $ & $0.875  $ & $ 125   \pm 17   $ \\
		$0.199  $ & $ 75    \pm 5    $ &  $0.4247 $ & $ 87.1  \pm 11.2 $ & $0.88   $ & $ 90    \pm 40   $ \\
		$0.20   $ & $ 72.9  \pm 29.6 $ &  $0.4497 $ & $ 92.8  \pm 12.9 $ & $0.9    $ & $ 117   \pm 23   $ \\
		$0.27   $ & $ 77    \pm 14   $ &  $0.4783 $ & $ 80.9  \pm 9    $ & $1.037  $ & $ 154   \pm 20   $ \\
		\hline
		\hline
	\end{tabular}
	\label{tab:h_clocks}
\end{table}

Then, the $\chi^2$ function of the cosmic chronometers is defined as

\begin{equation}\label{chi_H}
	\chi^{2}_{\text{CC}} = \sum_{i = 1}^{24}\Bigg(\frac{H_{\text{obs}}(z_i) - H_{\text{mod}}(z_i)}{\sigma_{H}^i}\Bigg)^2,
\end{equation}
where the $\sigma_i$ uncertainties in the $H(z)$ measurements for each data point $i$.

\begin{table*}[t]
	\renewcommand{\arraystretch}{2.0}
	\centering
	\medskip
	\caption{Confidence limits for the cosmological parameters using the BAO + CMB + CC + SNe Ia.
		The columns show the constraints on each model whereas the rows show each parameter considering in this analysis. In the last rows we have the Bayesian evidence, Bayes' factor and the interpretation.}
	\begin{tabular}{|c|c|c|c|c|c|c|c|}
		\hline
		Parameter &${\Lambda}$CDM & $\omega$CDM & Model \Rmnum{1}a & Model \Rmnum{1}b & Model \Rmnum{1}c &  Model \Rmnum{2} & Model \Rmnum{3}\\
		\hline
		\hline
		$h$ 	
		& $0.675\pm 0.005$ 
		& $0.679\pm 0.008$ 
		& $0.675\pm 0.006$
		& $0.681\pm 0.009$
		& $0.686\pm 0.009$
		& $0.679\pm 0.008$
		& $0.675\pm 0.005$
		\\
		$\Omega_{\text{b}}$	
		& $0.049\pm 0.001$ 
		& $0.048\pm 0.001$ 
		& $0.049\pm 0.001$
		& $0.048\pm 0.001$
		& $0.047\pm 0.001$
		& $0.048\pm 0.001$
		& $0.049\pm 0.001$
		\\
		$\Omega_{\text{dm}}$	
		& $0.267\pm 0.007$ 
		& $0.266\pm 0.007$ 
		& $0.267\pm 0.006$
		& $0.264\pm 0.007$
		& $0.267\pm 0.006$
		& $0.266\pm 0.007$
		& $0.266\pm 0.006$
		\\
		$\Omega_\text{k}$
		& $-$ 
		& $-$ 
		& $-$
		& $-$
		& $-0.053\pm 0.036$
		& $-$
		& $-$
		\\
		$\omega$
		& $-$ 
		& $-1.027\pm 0.040$ 
		& $-$
		& $-1.059\pm 0.059$
		& $-$
		& $-$
		& $-1.012 \pm 0.035$			
		\\
		$\tilde{\xi}$
		& $-$ 
		& $-$ 
		& $-0.0004\pm 0.008$
		& $-0.0097\pm 0.013$
		& $-0.026 \pm 0.020$
		& $ 0.012 \pm 0.019$
		& $-0.002 \pm 0.008$
		\\
		\hline \hline
		$\ln \mathcal{E}$
		& $-534.675 \pm 0.025$ 
		& $-537.491 \pm 0.008$ 
		& $-537.168 \pm 0.008$
		& $-539.232 \pm 0.006$
		& $-537.060 \pm 0.029$
		& $-536.199 \pm 0.007$
		& $ 535.902 \pm 0.599$
		\\
		$\ln B$
		& $-$ 
		& $-2.816 \pm 0.008$ 
		& $-2.493 \pm 0.008$
		& $-4.557 \pm 0.006$
		& $-2.385 \pm 0.029$
		& $-1.524 \pm 0.007$
		& $-1.227 \pm 0.599$
		\\
		Interpretation
		& $-$ 
		& Moderate
		& Weak
		& Moderate
		& Weak
		& Weak
		& Weak
		\\
		\hline
	\end{tabular}
	\label{tab:results-1}
\end{table*}

\begin{table*}[t]
	\renewcommand{\arraystretch}{2.0}
	\centering
	\medskip
	\caption{Confidence limits for the cosmological parameters using the BAO + CMB + CC + SNe Ia.
		The columns show the constraints on each extended model, with $q$ as free parameter, whereas the rows show each parameter considering in this analysis. In the last rows we have the Bayesian evidence, Bayes' factor and the interpretation.}
	\begin{tabular}{|c|c|c|c|c|c|c|c|}
		\hline
		Parameter &${\Lambda}$CDM & $\omega$CDM & Model \Rmnum{1}a & Model \Rmnum{1}b & Model \Rmnum{1}c &  Model \Rmnum{2} & Model \Rmnum{3}\\
		\hline
		\hline
		$h$ 	
	& $0.675^{+0.011}_{-0.011}$ 
	& $0.681^{+0.017}_{-0.017}$
	& $0.662^{+0.014}_{-0.015}$
	& $0.664^{+0.017}_{-0.017}$
	& $0.686^{+0.019}_{-0.018}$
	& $0.681^{+0.017}_{-0.017}$
	& $0.676^{+0.016}_{-0.017}$
	\\
	$\Omega_{\text{b}}$	
	& $0.049\pm 0.002$ 
	& $0.048^{+0.003}_{-0.002}$
	& $0.051^{+0.002}_{-0.002}$
	& $0.050^{+0.003}_{-0.002}$
	& $0.047^{+0.003}_{-0.002}$
	& $0.048^{+0.003}_{-0.002}$
	& $0.049^{+0.003}_{-0.002}$
	\\
	$\Omega_{\text{dm}}$	
	& $0.268^{+0.015}_{-0.014}$ 
	& $0.268^{+0.015}_{-0.014}$
	& $0.291^{+0.026}_{-0.024}$
	& $0.285^{+0.035}_{-0.032}$
	& $0.267^{+0.013}_{-0.012}$
	& $0.268^{+0.015}_{-0.014}$
	& $0.274^{+0.026}_{-0.025}$
	\\
	$\Omega_\text{k}$
	& $-$ 
	& $-$ 
	& $-$
	& $-$
	& $-0.052^{+0.067}_{-0.062}$
	& $-$
	& $-$
	\\
	$\omega$
	& $-$ 
	& $-1.051^{+0.095}_{-0.097}$ 
	& $-$
	& $-1.04^{+0.14}_{-0.15}$
	& $-$
	& $-$
	& $-1.07^{+0.14}_{-0.15}$			
	\\
	$ \tilde{\xi}$
	& $-$ 
	& $-$ 
	& $0.032^{+0.068}_{-0.066}$
	& $0.010\pm 0.10$
	& $-0.026^{+0.037}_{-0.036}$
	& $0.022^{+0.046}_{-0.046}$
	& $0.010^{+0.082}_{-0.089}$
	\\
	$q$
	& $0.998^{+0.013}_{-0.013}$
	& $0.993^{+0.016}_{-0.015}$
	& $0.977^{+0.045}_{-0.046}$
	& $0.986^{+0.055}_{-0.058}$
	& $1.10^{+0.28}_{-0.29}$
	& $0.994^{+0.016}_{-0.016}$
	& $0.975^{+0.061}_{-0.061}$
	\\
	\hline \hline
	$\ln \mathcal{E}$
	& $-543.921 \pm 0.007$ 
	& $-546.230 \pm 0.009$ 
	& $-543.831 \pm 0.055$
	& $-545.839 \pm 0.234$
	& $-541.955 \pm 0.069$
	& $-544.855 \pm 0.062$
	& $-545.852 \pm 0.058$
	\\
	$\ln B$
	& $-$ 
	& $-2.301 \pm 0.009$ 
	& $0.090  \pm 0.055$
	& $-1.918 \pm 0.234$
	& $1.966  \pm 0.069$
	& $-0.933 \pm 0.062$
	& $-1.231 \pm 0.058$
	\\
	Interpretation
	& $-$ 
	& Moderate
	& Inconclusive
	& Weak
	& Weak (favored)
	& Inconclusive
	& Weak
	\\
	\hline
	\end{tabular}
	\label{tab:results-2}
	\end{table*}


\subsection{Pantheon Type Ia Supernovae Sample}

The Type Ia Supernovae (SNe Ia) data is a relevant tool for understanding the actual evolution of the Universe. The Pantheon sample is the most recent SNe Ia sample, which consists of $1048$ measurements in the redshift range $0.01 < z < 2.3$ \cite{scolnic}. The observational distance moduli of SNe $\mu_{\text{obs}}$, can be calculated from

\begin{equation}\label{moduli}
	\mu_{\text{obs}} = m^{*}_{\text{B}} + \alpha X_{1} - \beta C - M_{\text{B}},
\end{equation}
where $m^{*}_{\text{B}}$, $X_{1}$ and $C$ are the B-band apparent magnitude, the stretch factor and color parameter, respectively. $M_{\text{B}}$ is the absolute magnitude. $\alpha$ and $\beta$ characterize the stretch and color-luminosity relationships, respectively. Commonly, $\alpha$ and $\beta$ are considered as free parameters and are constrained jointly with cosmological parameters. Nonetheless, this approach is model dependent, thus the distance calibrated by a cosmological could not be used to constrain parameters. To alleviate this problem, Ref.\cite{scolnic} proposes a method to calibrated SNe Ia named BEAMS with Bias Corrections (BBC) \cite{scolnic, marriner}. The Pantheon sample is calibrated using the BBC method, reducing the photometric calibration uncertainties (see more details in Refs.\cite{scolnic, marriner}). Then, to calculate the observational distance moduli we subtract $M_{\text{B}}$ from the apparent magnitude $m^{*}_{\text{B,corr}}$ and do not need the color and stretch corrections because now they are equal zero.
	
The theoretical distance modulus $\mu_{\text{th}}$ for a given supernova in redshift $z$ is expressed as

\begin{equation}
	\mu_{\text{th}} = 5\log_{10} \frac{d_L}{\text{Mpc}} + 25,
\end{equation}
where $d_{L} = (c/H_{0})D_{L}$ is the luminosity distance, with c is the speed of light, $H_{0}$ is the Hubble constant. Hubble-free luminosity distance is given by

\begin{equation}
	D_{L} = (1 + z_{\text{hel}})\int_{0}^{z_{\text{CMB}}} \frac{dz}{E(z)},
\end{equation}
where $E(z) = H(z)/H_0$, $z_{\text{CMB}}$ and $z_{\text{hel}}$ is the dimensionless Hubble parameter,  is the CMB frame and heliocentric redshift, respectively. From Eq. (\ref{moduli}) with $\alpha$ and $\beta$ equal zero, the observed distance moduli reads \cite{scolnic}

\begin{equation}
\mu_{\text{obs}} = m^{*}_{\text{B}} - \mathcal{M},
\end{equation}
with $m^{*}_{\text{B}}$ the B-band apparent magnitude and $\mathcal{M}$ is nuisance parameter that encompasses absolute magnitude $M_{\text{B}}$ and the Hubble constant $H_{0}$. The $\chi^2$ function from Pantheon data is given by

\begin{equation}\label{chi_pan}
\chi^{2}_{\text{Pan}} = \textbf{X}^{T}_{\text{Pan}}\cdot \textbf{C}^{-1}_{\text{Pan}} \cdot\textbf{X}_{\text{Pan}},
\end{equation}
where for the $i$-th SNe Ia, $\textbf{X}_{\text{Pan}} = \mu_{\text{obs,i}} - \mu_{\text{th,i}}$ and $\textbf{C}$ is the covariance matrix. We can rewrite Eq. (\ref{chi_pan}) as

\begin{equation}
\chi^{2}_{\text{Pan}} = \Delta\textbf{m}^{T}\cdot \textbf{C}^{-1} \cdot \Delta\textbf{m},
\end{equation}
with $ \Delta\textbf{m} = m_B - m_{\text{mod}}$, and

\begin{equation}
m_{\text{mod}} = 5\log_{10} D_L + \mathcal{M},
\end{equation}
in which $H_0$ in $d_L$ can be absorbed into $\mathcal{M}$, while the total covariance matrix $\textbf{C}$ is given by

\begin{equation}
\textbf{C} = \textbf{D}_{\text{stat}} + \textbf{C}_{\text{sys}},	
\end{equation}
where $\textbf{D}_{\text{stat}}$ is the diagonal covariance matrix of the statical uncertainties and $\textbf{C}_{\text{sys}}$, is the covariance matrix of systematics errors \cite{scolnic}. The nuisance parameter $\mathcal{M}$ could be marginalized following steps in Ref. \cite{conley}. Then, after the marginalization over $\mathcal{M}$, we define the following quantities

\begin{eqnarray}
a & = &  \Delta\textbf{m}^{T}\cdot \textbf{C}^{-1} \cdot \Delta\textbf{m},\\
b & = &  \Delta\textbf{m}^{T}\cdot \textbf{C}^{-1} \cdot \textbf{1}, \\
c & = &  \textbf{1}^{T} \cdot \textbf{C}^{-1} \cdot \textbf{1},
\end{eqnarray}
where $ \Delta\textbf{m} = m_B - m_{\text{mod}}$ and $\textbf{1}$ is a vector of unitary elements, finally, the $\chi^2$ function is reads

\begin{equation}
	\chi^{2}_{\text{Pan}} = a - \frac{b^2}{c} + \ln \frac{c}{2\pi}.
\end{equation}

For joint analysis, we consider the likelihood of each probe, namely $\mathcal{L}_{\text{joint}} = \mathcal{L}_{\text{BAO}} \times \mathcal{L}_{\text{CC}} \times \mathcal{L}_{\text{CMB}} \times \mathcal{L}_{\text{Pan}}$.

\begin{figure*}[!]
	\begin{center}
	\includegraphics[scale=0.50]{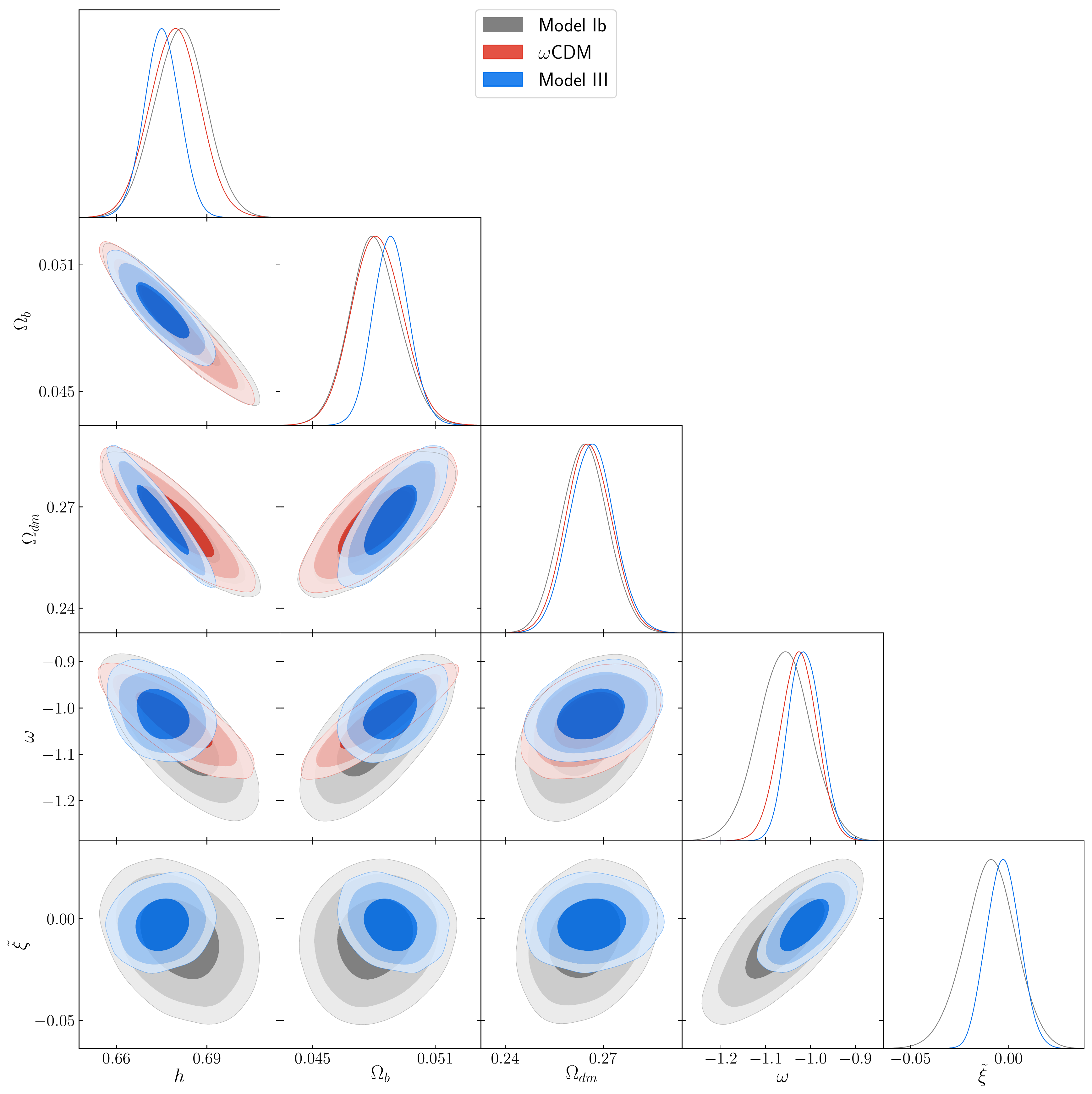}
	\caption{Confidence regions and PDFs for the parameters $h$, $\Omega_{\text{b}}$, $\Omega_{\text{dm}}$, $\Omega_{\text{k}}$, $\omega$ and $\tilde{\xi}$, for all the models studied considering combining data BAO + CMB + CC + SNe Ia.}
	\label{fig:regions1}
	\end{center}
\end{figure*}

\begin{figure*}[!]
	\begin{center}
\includegraphics[scale=0.50]{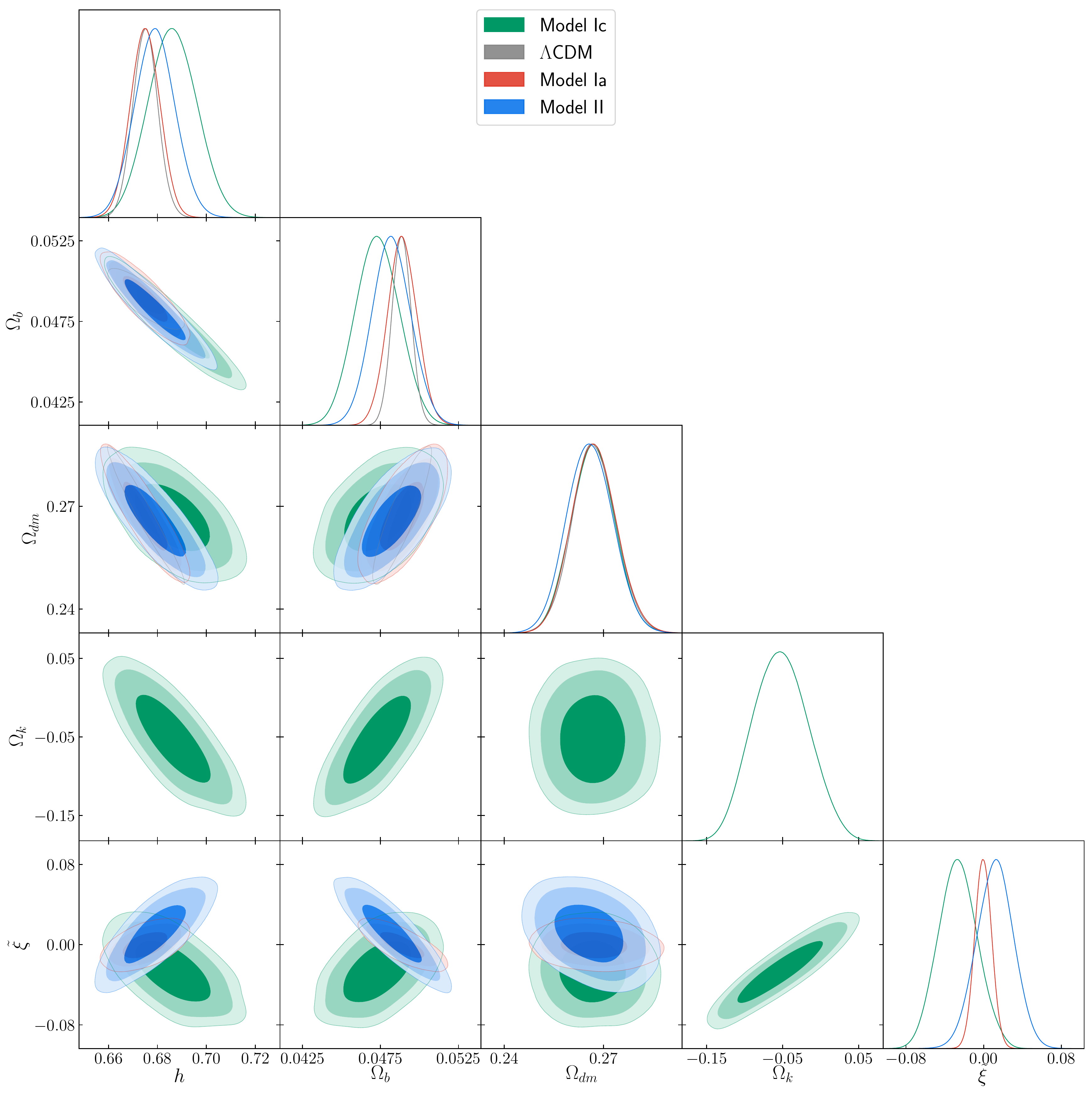}
\caption{Confidence regions and PDFs for the parameters $h$, $\Omega_{\text{b}}$, $\Omega_{\text{dm}}$, $\Omega_{\text{k}}$, $\omega$ and $\tilde{\xi}$, for all the models studied considering combining data BAO + CMB + CC + SNe Ia.}
	\label{fig:regions2}
	\end{center}
\end{figure*}

\section{Results and Conclusions}\label{sec:5}

In this work, we investigated some viscous dark energy models in the context of $\Lambda$CDM and the extend $\Lambda$CDM model. To analyze these models, we performed a Bayesian analysis in terms of the Jeffreys' Scale that evaluating the strength of evidence when comparing models \cite{trotta}. To achieve this analysis, we adopted the prior described in Table \ref{tab:prior} and considered distinct background data such as CMB priors distance, BAO measurements, cosmic chronometers, Pantheon Type Ia Supernova.

The main results of joint analysis (CMB + CC + BAO + SNe Ia) for $q$ fixed were summarized in Table \ref{tab:results-1}, including the mean and corresponding $1\sigma$ error of parameters for each model.  In the Figs. \ref{fig:regions1} and \ref{fig:regions2} show the posterior distributions and $1\sigma$, $2\sigma$ and $3\sigma$ contours regions for models studied. In the Table \ref{tab:results-1}, the dimensionless Hubble parameter converged for value obtained in the last Planck results \cite{planck, planck2018}. It is easy to see that $\Omega_{\text{b}}$ and $\Omega_{\text{dm}}$ were little affected by the test considered. For the Model \Rmnum{1}c, we found that the spatial curvature $\Omega_k = -0.053\pm 0.036$ was not compatible, $1\sigma$, with spatially flat Universe. For the Model \Rmnum{1}b and Model \Rmnum{3}, we got $\omega = - 1.059 \pm 0.059$ and $\omega = -1.012 \pm 0.035$ respectively, these results were still compatible with the standard cosmology ($\omega = -1$) but with very slightly preference for a phantom cosmology ($\omega < -1$) \cite{phantom}. In order to relieve the $H_0$ tension, we calculated the discrepancy between our results and Hubble constant local value \cite{riess2}. The values were $3.48\sigma$ for the Model \Rmnum{1}a, $2.92\sigma$ for the Model \Rmnum{1}b, $2.65\sigma$ for the Model \Rmnum{1}c, $3.11\sigma$ and $3.55\sigma$ for the Models \Rmnum{2} and \Rmnum{3}, respectively. We concluded that some models (excluding the Model \Rmnum{1}a) studied in this work alleviate the $H_0$ tension with emphasis on the Model \Rmnum{1}c, which has the lowest value of discrepancy.

\begin{figure*}[!]
	\begin{center}
		\includegraphics[scale=0.50]{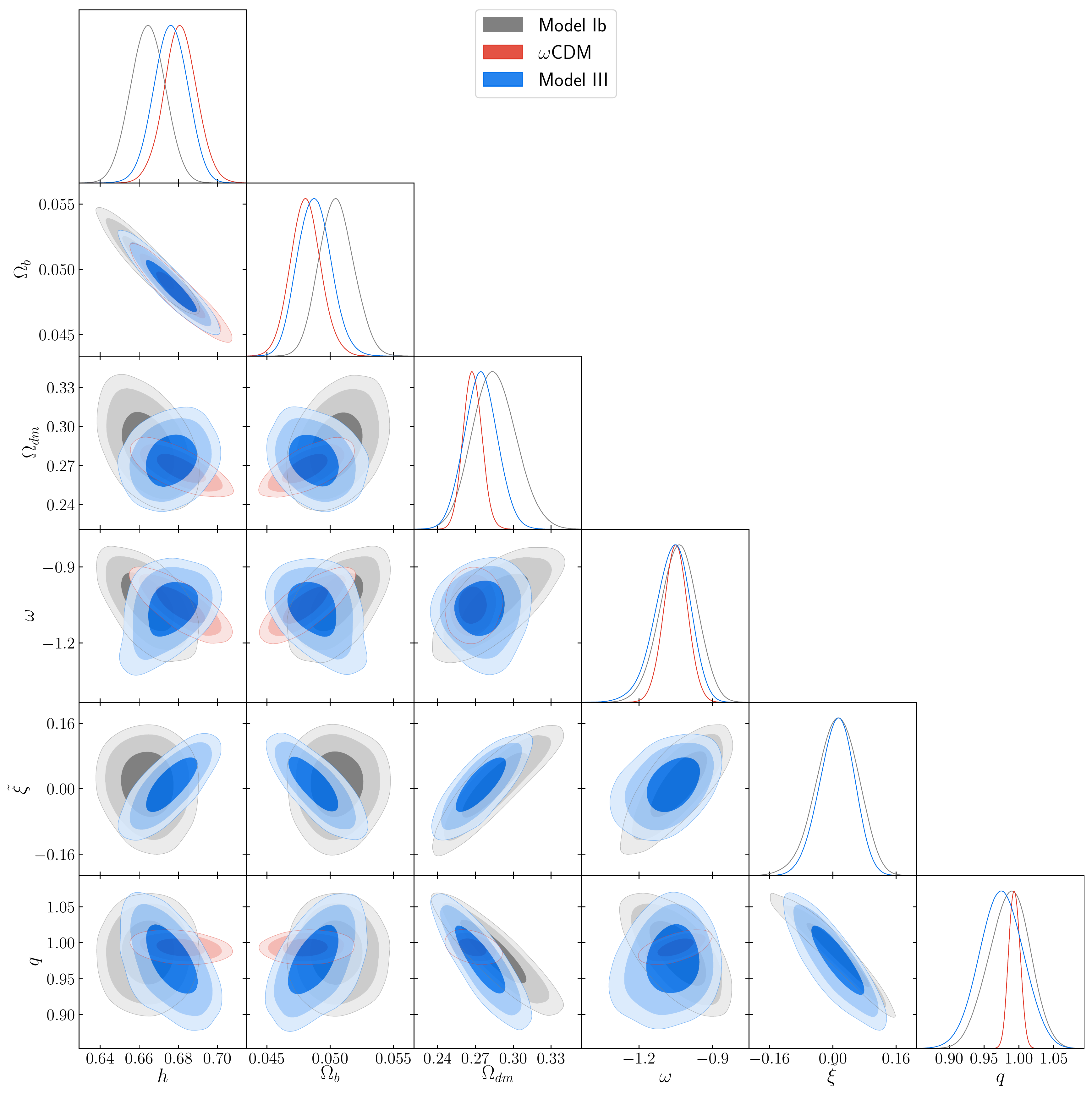}
		\caption{Confidence regions and PDFs for the parameters $h$, $\Omega_{\text{b}}$, $\Omega_{\text{dm}}$, $\Omega_{\text{k}}$, $\omega$, $ \tilde{\xi}$ and $q$, for all the models studied considering combining data BAO + CMB + CC + SNe Ia.}
		\label{fig:regions3}
	\end{center}
\end{figure*}

\begin{figure*}[!]
	\begin{center}
		\includegraphics[scale=0.50]{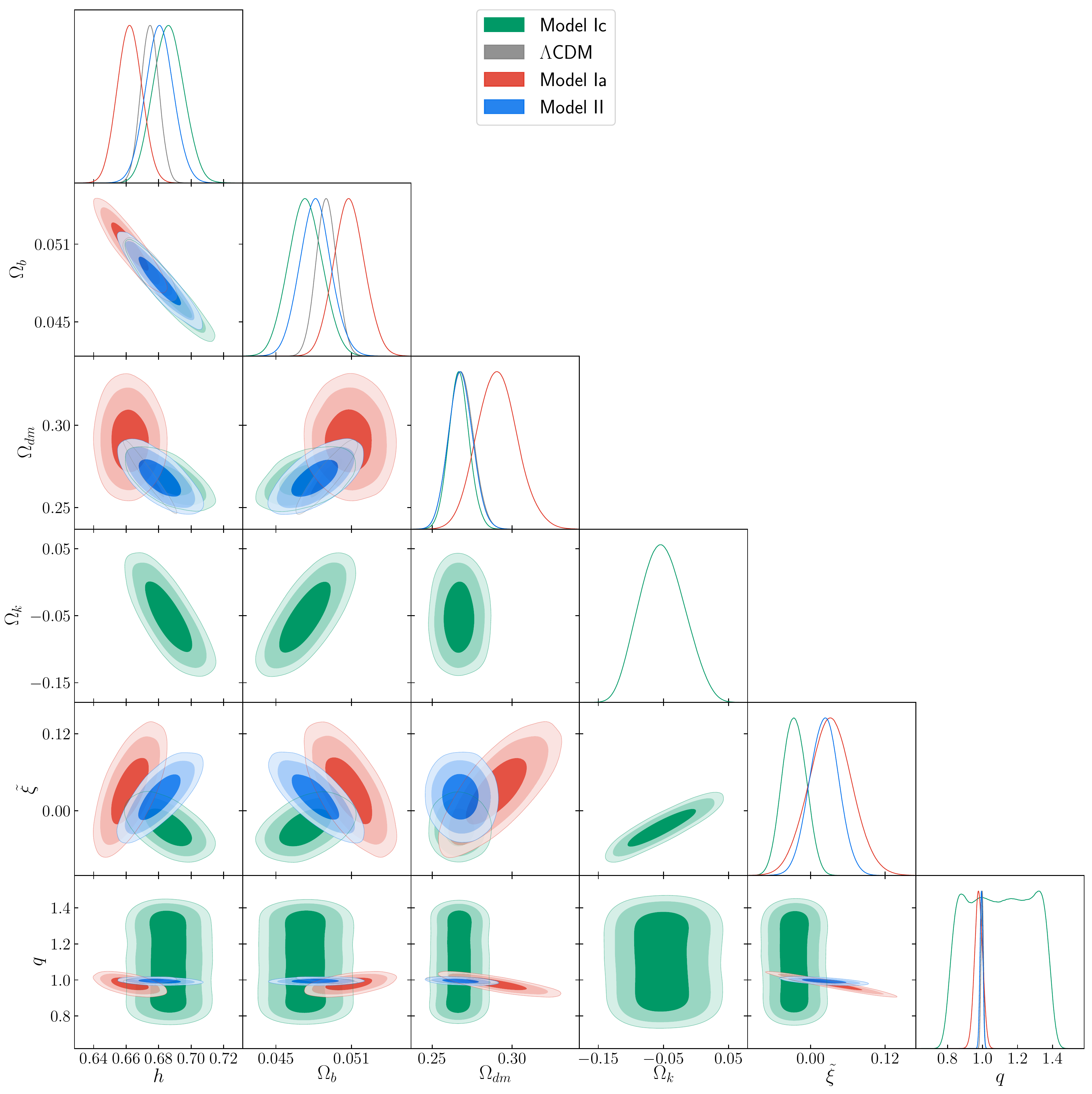}
		\caption{Confidence regions and PDFs for the parameters $h$, $\Omega_{\text{b}}$, $\Omega_{\text{dm}}$, $\Omega_{\text{k}}$, $\omega$, $\tilde{\xi}$ and $q$, for all the extended models studied considering combining data BAO + CMB + CC + SNe Ia.}
		\label{fig:regions4}
	\end{center}
\end{figure*}

The Figs. \ref{fig:regions3} and \ref{fig:regions4} showed the posterior distributions and $1\sigma$, $2\sigma$ and $3\sigma$ contours regions for extended models studied with $q$ as a free parameter. In the Table \ref{tab:results-2}, we showed that the values of $h$ obtained for Models \Rmnum{1}a and \Rmnum{1}b are lightly smaller than those obtained in the last Planck results. For Models \Rmnum{1}a and \Rmnum{1}b, with addition of the parameter $q$ in the analysis, the proportion of $\Omega_{\text{dm}}$ was slightly bigger than when $q$ was fixed. For the Model \Rmnum{1}c, we found that the results are compatible with first analysis ($q$ is fixed). We noted that by adding the parameter $q$ the value of bulk viscosity for the Models \Rmnum{1}a , \Rmnum{2} and \Rmnum{3} was  slightly increased.  Again, in order to relieve the $H_0$ tension, we calculated the discrepancy between our results and Hubble constant local value \cite{riess2}. The values were $5.29\sigma$ for the Model \Rmnum{1}a, $4.08\sigma$ for the Model \Rmnum{1}b, $2.47\sigma$ for the Model \Rmnum{1}c, $3.11\sigma$ and $3.64\sigma$ for the Models \Rmnum{2} and \Rmnum{3}, respectively. We concluded that models \Rmnum{1}a and \Rmnum{1}b not alleviate the $H_0$ tension, on the other hand, the Model \Rmnum{1}c has the lowest value of discrepancy.

For comparison of models, we calculated Bayes' factor considering  $\Lambda$CDM as the reference model. The values obtained for the logarithm of evidence, the logarithm of Bayes' factor and interpretation of Bayes' factor from the Jeffreys' scale were shown in Table \ref{tab:results-1}. By considering these data used, $\omega$CDM was disfavored with a moderate evidence in relation to the $\Lambda$CDM model. We also noticed that Model \Rmnum{1}b had an unfavorable moderately evidence, with $\ln B = -4.557 \pm 0.006$. Regarding the other models, for the Model \Rmnum{1}a, we obtained $\ln B = -2.493 \pm 0.008$, for Model \Rmnum{1}c, $\ln B = -2.385 \pm 0.029$, Model \Rmnum{2}, $\ln B = -1.524 \pm 0.007$ and for Model \Rmnum{3}, $\ln B = -1.227 \pm 0.559$, we found a disfavored weakly evidence. From the Bayesian comparison model analysis point of view and the data considered, we concluded that the viscous models studied in this work are ruled out.

Now, by considering the analysis with $q$ as a free parameter, the values obtained for the logarithm of evidence, the logarithm of Bayes' factor and interpretation of Bayes' factor from the Jeffreys' scale were shown in Table \ref{tab:results-2}. Then, from joint analysis, $\omega$CDM and Model \Rmnum{1}c were disfavored with a moderate and weak evidence, respectively, in relation to the $\Lambda$CDM model. The Models \Rmnum{1}a and \Rmnum{2} we can not made any conclusions about the evidence of this model in comparison to extended $\Lambda$CDM model. We found the positive logarithm of Bayes' factor ($\ln B = 1.966  \pm 0.069$) that indicated a weak evidence in favor of Model \Rmnum{1}c.

In summary, we showed that the viscous dark energy is compatible with the cosmological observations. However, the statistical constraints on the model parameters imply that the standard $\Lambda$CDM is recovered, i.e., bulk viscosity is zero. Moreover, we concluded from Bayesian analysis standpoint that our model has disfavored moderate and weak evidence compared with $\Lambda$CDM. We concluded that $\Lambda$CDM still has the best efficiency to explain the data used in this work; this conclusion is dependent on either by analyzing the parameters, or the Bayesian evidence.

Finally, it is worth emphasizing that in order to obtain a robust formulation (theoretical and observational), we need to investigate both background expansion and perturbations effects. This issue will be investigated in the future work.

\begin{acknowledgments}
The authors thank Brazilian scientific and financial support federal agencies, CAPES and CNPq. RS thanks CNPq (Grant No. 303613/2015-7) for financial support. This work was supported by High-Performance Computing Center (NPAD)/UFRN. W. J. C. da Silva thanks Antonella Cid for her assistance in the Bayesian analysis and Jailson Alcaniz for discussions and support in Rio de Janeiro. W. J. C. da Silva thanks Observat\'orio Nacional (ON) for hospitality during the development of this work. We thanks G.M. Viswanathan for feedback.
\end{acknowledgments}

\end{document}